\documentclass{article}
\usepackage{amsmath}

\numberwithin{equation}{section}
\input{tcilatex}
\begin{document}


\centerline{\large\bf Stochastic local operations and classical
communication} 
\centerline{\large\bf properties of the n-qubit symmetric Dicke
states \footnote{The paper was supported by NSFC(Grants No.
60433050, 60673034, and 10875061)  }} 

\centerline{Dafa Li$^{a}$\footnote{email
address:dli@math.tsinghua.edu.cn}, Xinxin Li$^{b}$  , Hongtao
Huang$^{c}$,  Xiangrong Li$^{d}$}

\centerline{$^a$ Dept of mathematical sciences, Tsinghua University,
Beijing 100084 CHINA}

\centerline{$^b$ Dept. of Computer Science, Wayne State University,
Detroit, MI 48202, USA}

\centerline{$^c$ Electrical Engineering and Computer Science Department} %
\centerline{ University of Michigan, Ann Arbor, MI 48109, USA}

\centerline{$^d$ Department of Mathematics, University of
California, Irvine, CA 92697-3875, USA}



Abstract

Recently, several schemes for the experimental creation of Dicke states were
described. In this paper, we show that all the $n$-qubit symmetric Dicke
states with $l$ ($2\leq l\leq (n-2)$) excitations are inequivalent to the $%
|GHZ\rangle $ state or the $|W\rangle $ state under SLOCC, that the even $n$%
-qubit symmetric Dicke state with $n/2$ excitations is inequivalent to any
even $n$-qubit symmetric Dicke state with $l\neq n/2$ excitations under
SLOCC, and that all the $n$-qubit symmetric\ Dicke states with $l$ ($2\leq
l\leq (n-2)$) excitations satisfy Coffman, Kundu and Wootters' generalized
monogamy inequality $C_{12}^{2}+...+C_{1n}^{2}<C_{1(2...n)}^{2}<1$.

Pacs: 03.65.Ud, 03.67.Mn.

Keywords: CKW's monogamy inequality, concurrence, SLOCC entanglement
classification, symmetric Dicke states.

\section{Introduction}

Stockton et al. pointed out that the symmetric systems are experimentally
interesting because it is easier to nonselectively address an entire
ensemble of particles rather than individually address each member \cite%
{Stockton}. Due to the symmetry under permutations of the qubits, the
symmetric Dicke states become important. Dicke states are considered to be
the simultaneous eigenstates of both the square of the total spin operator $%
\hat{S}^{2}$ and its $z$-component $\hat{S}_{z}$ \cite{Thiel}. In \cite%
{Stockton}, the $n$-qubit symmetric Dicke states with $l$ excitations, where 
$1\leq l\leq (n-1)$, were defined as follows.

\begin{equation}
|l,n\rangle =\sum_{i}P_{i}|1_{1}1_{2}...1_{l}0_{l+1}...0_{n}\rangle ,
\label{Dicke}
\end{equation}%
where $\{P_{i}\}$ is the set of all the distinct permutations of the qubits.
But, from the definition in Eq. (\ref{Dicke}),\ it is not obvious what
position each \textquotedblleft 1\textquotedblright\ or \textquotedblleft $0$%
\textquotedblright\ occurs in $P_{i}|1_{1}1_{2}...1_{l}0_{l+1}...0_{n}%
\rangle $. For our further consideration it is desirable to rewrite Eq. (\ref%
{Dicke}) by using a binary basis in the following form:

\begin{equation}
|l,n\rangle =(1/\sqrt{\left( _{l}^{n}\right) })\sum_{0\leq
i_{1}<i_{2}<...i_{l}\leq n-1}|2^{i_{1}}+2^{i_{2}}+...+2^{i_{l}}\rangle
,1\leq l\leq n-1.  \label{def1}
\end{equation}%
It is easy to see that just $l$ ones occur in each basis term of the state $%
|l,n\rangle $ and $|l,n\rangle $ is symmetric under permutations of the
qubits. $|1,n\rangle $ is just the $|W\rangle $ state, i.e., $|1,n\rangle
=(|100...0\rangle +|01...0\rangle +...+|0...01\rangle )/\sqrt{n}$.

Several schemes for the experimental creation of Dicke states were described
in \cite{Thiel, Unanyan, Duan, Retzker, Stockton04, Lopez, Lemr, Scully}.
For example, the authors in \cite{Bourennane, Eibl, Mikami} realized the $%
|W\rangle $ state of three qubits in a photonic system, while the $|W\rangle 
$ state of eight qubits was created with trapped ions in \cite{Haffner}. The
authors in \cite{Kiesel} generated a four-qubit symmetric Dicke state with
two excitations and discussed its application in quantum communication. The
authors in \cite{Thiel}\cite{Lopez} reported the realistic proposals to
generate Dicke states in specific physical systems. The cavity QED schemes
for generating symmetric Dicke states were discussed in \cite{Zheng08}. The
methods for detecting entanglement around symmetric Dicke states were
presented in \cite{Geza}.

Quantum entanglement is a quantum mechanical resource and plays a key role
in quantum computation and quantum information. Recently, many authors have
exploited SLOCC (stochastic local operations and classical communication)
entanglement classification \cite{Bennett, Dur, Moor2, Miyake, Lamata06,
Lamata07, LDF07a, LDF07b}. If two states can be obtained from each other by
means of local operations and classical communication (LOCC) with nonzero
probability, we say that two states have the same kind of entanglement\cite%
{Bennett}. In \cite{Dur},\ D\"{u}r et al. showed that for pure states of
three qubits there are six inequivalent SLOCC entanglement classes, of which
two are true entanglement classes: $|GHZ\rangle $ and $|W\rangle $.
Verstraete et al. \cite{Moor2} discussed the entanglement classes of four
qubits under SLOCC and pointed out there exist nine families of states
corresponding to nine different ways of entangling four qubits.

As indicated in \cite{Dur}, if two states are SLOCC equivalent, then they
are suited to do the same tasks of QIT. In this paper, by means of the SLOCC
invariant for $n$ qubits \cite{LDF07a}, we investigate the SLOCC\ properties
of the $n$-qubit symmetric Dicke states with $l$ excitations. The different
SLOCC invariants were proposed in \cite{Miyake}\cite{Sudbery}. For the
readability, we list the SLOCC invariant for $n$ qubits \cite{LDF07a} as
follows.

Let $|\psi \rangle $\ and $|\psi ^{\prime }\rangle $\ be any states of $n$\
qubits. Then we can write $|\psi \rangle =\sum_{i=0}^{2^{n}-1}a_{i}|i\rangle 
$ and $|\psi ^{\prime }\rangle =\sum_{i=0}^{2^{n}-1}a_{i}^{\prime }|i\rangle 
$. It is well known from \cite{Dur}\ that $|\psi \rangle $ is equivalent to $%
|\psi ^{\prime }\rangle $ under SLOCC if and only if 
\begin{equation}
|\psi \rangle =\underbrace{\alpha \otimes \beta \otimes \gamma \cdots }%
_{n}|\psi ^{\prime }\rangle ,  \label{slocc-eq}
\end{equation}%
where $\alpha $, $\beta $, $\gamma $, ... are invertible local operators.

For even $n$ qubits, if $|\psi \rangle $ is equivalent to $|\psi ^{\prime
}\rangle $ under SLOCC then $|\psi \rangle $\ and $|\psi ^{\prime }\rangle $
satisfy the following equation \cite{LDF07a}:

\begin{equation}
\tau (\psi )=\tau (\psi ^{\prime })\underbrace{|\det (\alpha )\det (\beta
)\det (\gamma )...|}_{n},  \label{evenre2}
\end{equation}
\ \ \ \ where\ \ \ 

\begin{eqnarray}
&&\tau (\psi )=2|\mathcal{I}^{\ast }(a,n)|\text{,}  \notag \\
\mathcal{I}^{\ast }(a,n) &=&\sum_{i=0}^{2^{n-2}-1}sgn^{\ast
}(n,i)(a_{2i}a_{(2^{n}-1)-2i}-a_{2i+1}a_{(2^{n}-2)-2i}),  \label{evenredua}
\end{eqnarray}%
in which

\begin{equation*}
sgn^{\ast }(n,i)=%
\begin{cases}
(-1)^{N(i)} & \text{for }0\leq i\leq 2^{n-3}-1\text{,}\qquad  \\ 
(-1)^{n+N(i)} & \text{for }2^{n-3}\leq i\leq 2^{n-2}-1\text{.}%
\end{cases}%
\end{equation*}%
The $N(i)$\ above is defined as follows. Let $i_{n-1}...i_{1}i_{0}$\ be an $%
n-$bit binary representation of $i$. That is, $%
i=i_{n-1}2^{n-1}+...+i_{1}2^{1}+i_{0}2^{0}$. Then, let $N(i)$\ be the number
of the occurrences of \textquotedblleft $1$\textquotedblright\ in $%
i_{n-1}...i_{1}i_{0}$. Note that when $n$\ is even, $%
(-1)^{n+N(i)}=(-1)^{N(i)}$, and $sgn^{\ast }(n,i)=$\ $(-1)^{N(i)}$\ for $%
0\leq i\leq 2^{n-2}-1$. Thus, Eq. (\ref{evenredua}) can be simplified as Eq.
(\ref{newdef}) in this paper. 

For even $n$ qubits, we can define $\tau (\psi ^{\prime })$ in Eq. (\ref%
{evenre2}) from the definition of $\tau (\psi )$ in Eq. (\ref{evenredua})\
by replacing the amplitudes $a_{k}$ of $|\psi \rangle $ with the amplitudes $%
a_{k}^{\prime }$ of $|\psi ^{\prime }\rangle $.

For odd $n$ qubits, if $|\psi \rangle $ is equivalent to $|\psi ^{\prime
}\rangle $ under SLOCC then $|\psi \rangle $\ and $|\psi ^{\prime }\rangle $
satisfy the following equation \cite{LDF07a}:

\begin{equation}
\tau (\psi )=\tau (\psi ^{\prime })\underbrace{|\det^{2}(\alpha
)\det^{2}(\beta )\det^{2}(\gamma )...|}_{n},  \label{oddre2}
\end{equation}%
where

\begin{equation}
\tau (\psi )=4|(\overline{\mathcal{I}}(a,n))^{2}-4\mathcal{I}^{\ast }(a,n-1)%
\mathcal{I}_{+2^{n-1}}^{\ast }(a,n-1)|,  \label{odd-residua-def}
\end{equation}%
in which

\begin{eqnarray}
&&\overline{\mathcal{I}}(a,n)=  \notag \\
&&%
\sum_{i=0}^{2^{n-3}-1}(-1)^{N(i)}[(a_{2i}a_{(2^{n}-1)-2i}-a_{2i+1}a_{(2^{n}-2)-2i})
\notag \\
&&-(a_{(2^{n-1}-2)-2i}a_{(2^{n-1}+1)+2i}-a_{(2^{n-1}-1)-2i}a_{2^{n-1}+2i})],
\label{odd-def-2}
\end{eqnarray}%
and

\begin{eqnarray}
&&\mathcal{I}_{+2^{n-1}}^{\ast }(a,n-1)=\sum_{i=0}^{2^{n-3}-1}sgn^{\ast
}(n-1,i)\times  \notag \\
&&(a_{2^{n-1}+2i}a_{(2^{n}-1)-2i}-a_{2^{n-1}+1+2i}a_{(2^{n}-2)-2i}),
\label{odd-def-3}
\end{eqnarray}

For odd $n$ qubits, we can also define $\tau (\psi ^{\prime })$ in Eq. (\ref%
{oddre2})\ from the definition of $\tau (\psi )$ in Eqs. (\ref%
{odd-residua-def}), \ref{odd-def-2}\ and \ref{odd-def-3}\ by replacing the
amplitudes $a_{k}$ of $|\psi \rangle $ with the amplitudes $a_{k}^{\prime }$
of $|\psi ^{\prime }\rangle $. \ 

By Eqs. (\ref{evenre2}) and (\ref{oddre2}), we obtain the following
corollary 1.

Corollary 1. For any $n$ qubits, if $\tau (\psi )=0$ but $\tau (\psi
^{\prime })\neq 0$ or $\tau (\psi )\neq 0$ but $\tau (\psi ^{\prime })=0$,
then $|\psi \rangle $\ and $|\psi ^{\prime }\rangle $ are inequivalent under
SLOCC.

A simple calculation shows that for any $n$-qubit $|GHZ\rangle =$ $%
(|0\rangle ^{\otimes n}+|1\rangle ^{\otimes n})/\sqrt{2}$, $\tau (GHZ)=1$,
and for any $n$-qubit $|W\rangle $, $\tau (W)=0$. By Corollary 1, the states 
$|GHZ\rangle $ and $|W\rangle $ of any $n$\ qubits are inequivalent under
SLOCC.

This paper is organized as follows. In Sec. 2, we show that all the Dicke
states with $l$ excitations\textbf{\ }are true entangled. In Sec. 3, we
discuss the SLOCC\ properties of the even $n$-qubit symmetric Dicke states.
In Sec. 4, we analyze the SLOCC\ properties of the odd $n$-qubit symmetric
Dicke states. In Sec. 5, we discuss the monogamy inequality for Dicke states.

\ \ \ \ \ \ \ \ \ \ \ \ \ \ \ \ \ \ \ \ \ \ \ \ \ \ \ \ \ \ \ \ \ \ \ \ \ \
\ \ \ \ \ \ \ \ \ \ \ \ \ \ \ \ \ \ \ \ \ \ \ \ \ \ \ \ \ \ \ \ \ \ \ \ \ \
\ \ \ \ \ \ \ \ \ \ \ \ \ \ \ \ \ \ \ \ \ \ \ \ \ \ \ \ \ \ \ \ \ \ \ \ \ \ 

\section{\textbf{\ }All the Dicke states with $l$ excitations\textbf{\ are
true entangled}}

When $l=1$ and $n=2$, it reduces to Bell state. When $l=1$ and $n\geq 3$, $%
|l,n\rangle $ is the $|W\rangle $ state. Let us consider $l>1$ and $n\geq 3$%
. Assume that $|l,n\rangle $ is a product state. Then, we can write $%
|l,n\rangle $ $=|\phi \rangle \otimes |\omega \rangle $, where $|\phi
\rangle $ is a state of $k$ qubits and $|\omega \rangle $ is a state of $%
(n-k)$ qubits.

Case 1. If $l$ ones occur in some basis term of $|\phi \rangle $, then only
zeros occur in each basis term of $|\omega \rangle $. It is impossible.

Case 2. If only zeros occur in some basis term of $|\phi \rangle $, then $l$
ones must occur in each basis term of $|\omega \rangle $. It is also
impossible.

Case 3. If $t$ ones, where $1\leq t\leq l-1$, occur in some basis term of $%
|\phi \rangle $, then each basis term of $|\omega \rangle $ must contain $%
(l-t)$ ones. It implies that each basis term of $|\phi \rangle $ must
contain $t$ ones. Thus, each basis term of $|\phi \rangle $ contains $t$
ones and $|\phi \rangle $ has $\left( _{t}^{k}\right) $ terms,\ while each
basis term of $|\omega \rangle $ contains $(l-t)$ ones and $|\omega \rangle $
has $\left( _{l-t}^{n-k}\right) $ terms. Thus, $|\phi \rangle \otimes
|\omega \rangle $ has $\left( _{t}^{k}\right) \left( _{l-t}^{n-k}\right) $
terms. However, $|l,n\rangle $ has $\left( _{l}^{n}\right) $ terms. Hence,
this case is impossible.

From the above cases, clearly $|l,n\rangle $ \ ($1\leq l\leq (n-1)$) is a
true entangled state, i.e., not a product state.

For example, for $n$ qubits, $|2,n\rangle =(1/\sqrt{n(n-1)/2})\sum_{0\leq
i<j\leq n-1}|2^{i}+2^{j}\rangle $. It means that $(n-2)$ zeros and two ones
occur in each term of the state $|2,n\rangle $. So, $|2,n\rangle $ has $%
n(n-1)/2$ terms.\ For example, $|2,4\rangle =(|0011\rangle +|0101\rangle
+|0110\rangle +|1001\rangle +|1010\rangle +|1100\rangle )/\sqrt{6}$. It was
proven that $|2,4\rangle $ is a true entangled state in \cite{LDF07b}.

\section{Even $n$-qubit symmetric Dicke states}

In this section, we show that the even $n$-qubit symmetric Dicke state with $%
n/2$ excitations is inequivalent to any even $n$-qubit symmetric Dicke state
with $l\neq n/2$ excitations under SLOCC, and that the even $n$-qubit
symmetric Dicke states with $l$ excitations, where $2\leq l\leq (n-2)$, are
different from the even $n$-qubit $|GHZ\rangle $ and $|W\rangle $ states
under SLOCC, respectively.

Note that the states $|l,n\rangle $ and $|(n-l),n\rangle $ are equivalent
under SLOCC by Lemma 1 in Appendix A. Hence, it is enough to consider $2\leq
l\leq n/2$.

\subsection{Dicke state with $n/2$ excitations is inequivalent to Dicke
state with $l\neq n/2$ excitations under SLOCC.}

\subsubsection{$\protect\tau (|n/2,n\rangle )=1$ for Dicke state with $n/2$
excitations}

By the definition in Eq. (\ref{def1}), for the state $|l,n\rangle $ ,\ the
amplitudes $a_{2^{i_{1}}+2^{i_{2}}+...+2^{i_{l}}}=1/\sqrt{\left(
_{l}^{n}\right) }$. Otherwise, $a_{k}=0$. In order to compute $\tau (\psi )$%
, by Remark 2.1 in \cite{LDF07e} $\tau (\psi )$ in Eq. (\ref{evenredua}) can
be rewritten as\ 

\begin{equation}
\tau (\psi )=2|\sum_{k=0}^{2^{n-1}-1}(-1)^{N(k)}a_{k}a_{(2^{n}-1)-k}|.
\label{newdef}
\end{equation}%
In Eq. (\ref{newdef}), each term is of the form $%
(-1)^{N(k)}a_{k}a_{(2^{n}-1)-k}$. Note that for the indexes, $%
k+(2^{n}-1)-k=(2^{n}-1)$, whose binary number is $\underbrace{11...1}_{n}$.
For the state $|n/2,n\rangle $, we only consider the term

$%
a_{2^{i_{1}}+2^{i_{2}}+...+2^{i_{(n/2)}}}a_{2^{j_{1}}+2^{j_{2}}+...+2^{j_{(n/2)}}} 
$, where $0\leq i_{1}<i_{2}<...<i_{(n/2)}\leq n-1$ and $0\leq
j_{1}<j_{2}<...<j_{(n/2)}\leq n-1$, and $%
2^{i_{1}}+2^{i_{2}}+...+2^{i_{(n/2)}}+2^{j_{1}}+2^{j_{2}}+...+2^{j_{(n/2)}}=2^{n}-1 
$. Clearly, $%
a_{2^{i_{1}}+2^{i_{2}}+...+2^{i_{(n/2)}}}a_{2^{j_{1}}+2^{j_{2}}+...+2^{j_{(n/2)}}}=1/\left( _{n/2}^{n}\right) 
$. Note that $(-1)^{N(2^{i_{1}}+2^{i_{2}}+...+2^{i_{(n/2)}})}=(-1)^{n/2}$.
Clearly, there are $\left( _{n/2}^{n}\right) /2$ terms being of the form $%
a_{2^{i_{1}}+2^{i_{2}}+...+2^{i_{(n/2)}}}a_{2^{j_{1}}+2^{j_{2}}+...+2^{j_{(n/2)}}} 
$, and each term has the same sign $(-1)^{n/2}$. Hence, by Eq. (\ref{newdef}%
), $\tau (|n/2,n\rangle )=1$. For example, $\tau (|2,4\rangle )=1$ and $\tau
(|3,6\rangle )=1$.

\subsubsection{$\protect\tau (|l,n\rangle $ $)=0$ for Dicke state with $%
1\leq l\leq (n-1)$ but $l\neq n/2$}

Let us prove that each term $a_{k}a_{(2^{n}-1)-k}$ in Eq. (\ref{newdef})
vanishes\textbf{\ }when \textbf{\ }$l\neq n/2$. By the definition in Eq. (%
\ref{def1}),\ if $N(k)\neq l$, then $a_{k}=0$. Otherwise, $N(k)=l$ and $%
a_{k}=1/\sqrt{\left( _{i}^{n}\right) }$. However, when $N(k)=l$, $%
N((2^{n}-1)-k)=n-l\neq l$. Thus, $a_{(2^{n}-1)-k}=0$. Therefore, $\tau
(|l,n\rangle $ $)=0$. Especially, $\tau (W)=0$.

From the above discussion, $\tau (|n/2,n\rangle )=1$ while $\tau
(|l,n\rangle $ $)$ $=0$, where $1\leq l\leq (n-1)$ but $l\neq n/2$. By
Corollary 1, the state $|n/2,n\rangle $ is different from the states $%
|l,n\rangle $ ($1\leq l\leq (n-1)$ but $l\neq n/2$) under SLOCC.

\subsection{All the $n$-qubit symmetric Dicke states are inequivalent to the 
$n$-qubit $|GHZ\rangle $ state under SLOCC.}

In order to show that the states $|l,n\rangle $ ($2\leq l\leq (n-2)$) are
different from the $|GHZ\rangle $ and $|W\rangle $ states under SLOCC
respectively, let us consider the following quantity for any state $|\Omega
\rangle =\sum_{k=0}^{2^{n}-1}b_{k}|k\rangle $,

\begin{eqnarray}
D^{(l)}(\Omega ) &=&(b_{1+\Delta }b_{4+\Delta }-b_{0+\Delta }b_{5+\Delta
})(b_{11+\Delta }b_{14+\Delta }-b_{10+\Delta }b_{15+\Delta })  \notag \\
&&-(b_{3+\Delta }b_{6+\Delta }-b_{2+\Delta }b_{7+\Delta })(b_{9+\Delta
}b_{12+\Delta }-b_{8+\Delta }b_{13+\Delta }),  \label{D-cretiria}
\end{eqnarray}%
where 
\begin{equation*}
\Delta =\left\{ 
\begin{array}{rcc}
0 & : & l=2\text{,} \\ 
2^{4}+2^{5}+...+2^{l+1} & : & l\geq 3\text{.}%
\end{array}%
\right.
\end{equation*}

Lemma 2 in Appendix A says that for $n$ qubits, if $|\psi \rangle $ is
equivalent to $|GHZ\rangle $ under SLOCC then $D^{(l)}(\psi )=0$, where $%
2\leq l\leq (n-2)$. However, Lemma 4 in Appendix A says that for states $%
|l,n\rangle $ , $D^{(l)}(|l,n\rangle $ $)\neq 0$, where $2\leq l\leq (n-2)$.
Therefore, the states $|l,n\rangle $ ($2\leq l\leq (n-2)$) are different
from the $|GHZ\rangle $ state under SLOCC. By means of Corollary 1, we can
also verify that the states $|l,n\rangle $ ($2\leq l\leq (n-1)$ but $l\neq
n/2$) are different from the $|GHZ\rangle $ state under SLOCC. This is
because that $\tau (|l,n\rangle $ $)=0$ while $\tau (GHZ)=1$ for even $n$
qubits. See Sec. 3.1.

Remark 1. $\tau (\psi )$ in Eq. (\ref{evenredua}) is considered as the
residual entanglement for even $n$ qubits in \cite{LDF07a} and \cite{LDF07e}%
. From the above discussion, we can say that for even $n$\ qubits, the
states $|n/2,n\rangle $ and $|GHZ\rangle $ possess the maximal residual
entanglement $\tau =1$ while the residual entanglement $\tau $ for the Dicke
states $|l,n\rangle $ ($l\neq n/2$) vanishes.

\subsection{All the $n$-qubit symmetric Dicke states with $2\leq l\leq (n-2)$
excitations are different from the $n$-qubit $|W\rangle $ state under SLOCC.}

Lemma 3 in Appendix A says that for $n$ qubits, if $|\psi \rangle $ is
equivalent to $|W\rangle $ under SLOCC then $D^{(l)}(\psi )=0$, where $2\leq
l\leq (n-2)$. However, Lemma 4 in Appendix A says for states $|l,n\rangle $
, $D^{(l)}(|l,n\rangle $ $)\neq 0$, where $2\leq l\leq (n-2)$. Therefore,
the states $|l,n\rangle $ ($2\leq l\leq (n-2)$) are different from the $%
|W\rangle $ state under SLOCC. By means of Corollary 1, we can also verify
that the state $|n/2,n\rangle $ is different from the $|W\rangle $ state
under SLOCC because $\tau (|n/2,n\rangle )=1$ and $\tau (W)$ $=0$.

Conjecture. For even $n$ qubits, perhaps $|2,n\rangle $, $|3,n\rangle $, ...
, and $|(n/2-1),n\rangle $ are different from each other under SLOCC because 
$D^{(k)}(|l,n\rangle $ $)=0$ whenever $k\neq l$\ while $D^{(l)}(|l,n\rangle $
$)\neq 0$.

\section{Odd $n$-qubit symmetric Dicke states}

In this section, we demonstrate that the odd $n$-qubit Dicke states with $l$
excitations, where $2\leq l\leq (n-2)$, are different from the odd $n$-qubit$%
\ |GHZ\rangle $ and $|W\rangle $\ states under SLOCC, respectively.

Note that the states $|l,n\rangle $ and $|(n-l),n\rangle $ are equivalent
under SLOCC by Lemma 1 in Appendix A. Hence, it is enough to consider that ($%
2\leq l\leq (n-1)/2$) in this section.

By means of Corollary 1, we demonstrate that for odd $n$ qubits, the states $%
|l,n\rangle $ ($2\leq l<(n-1)/2$) are inequivalent to the $|GHZ\rangle $
state under SLOCC below.

\subsection{All the $n$-qubit symmetric Dicke states are inequivalent to the 
$n$-qubit $|GHZ\rangle $ state under SLOCC$.$}

First let us prove that $\tau (|l,n\rangle $ $)=0$ when $1\leq l\leq (n-1)/2$
as follows.

Note that each term in Eq. (\ref{odd-def-3}) is of the form $%
a_{2^{n-1}+k}a_{(2^{n}-1)-k}$. For the state $|l,n\rangle $ ,\ we want to
show $a_{2^{n-1}+k}a_{(2^{n}-1)-k}=0$.\ Note that $%
2^{n-1}+k+(2^{n}-1)-k=2^{n}-1+2^{n-1}$, whose binary number is $10%
\underbrace{11...1}_{n-1}$. That is, $N(2^{n}-1+2^{n-1})=n$. Since $2l\leq
(n-1)$, it is impossible that $N(2^{n-1}+k)=N((2^{n}-1)-k)=l$. It says that $%
a_{2^{n-1}+k}a_{(2^{n}-1)-k}=0$. Thus, $\mathcal{I}_{+2^{n-1}}^{\ast
}(a,n-1)=0$.

Note that each term of $\overline{\mathcal{I}}(a,n)$ in Eq. (\ref{odd-def-2}%
) is of the form $a_{k}a_{(2^{n}-1)-k}$. It is trivial that $%
k+(2^{n}-1)-k=(2^{n}-1)$ and $N(2^{n}-1)=n$. As discussed above, it is
impossible that $N(k)=N((2^{n}-1)-k)=l$ because $2l\leq (n-1)$. Hence, $%
a_{k}a_{(2^{n}-1)-k}=0$. Also, $\overline{\mathcal{I}}(a,n)=0$.

From the discussion above, $\tau (|l,n\rangle $ $)=0$ when $1\leq l\leq
(n-1)/2$. For example, by calculating it is easy to see that $\tau
(|2,5\rangle )=0$.

Since $\tau (GHZ)=1$ while $\tau (|l,n\rangle )=0$ ($1\leq l\leq (n-1)/2$),
by Corollary 1 the states $|l,n\rangle $ \ ($1\leq l\leq (n-1)/2$) are
different from the $|GHZ\rangle $ state under SLOCC. Lemmas 2 and 4 in
Appendix A also verify this fact. Lemma 2 says that $D^{(l)}(\psi )=0$ for
any state $|\psi \rangle $ in the $|GHZ\rangle $ class while Lemma 4 says
that $D^{(l)}(|l,n\rangle )\neq 0$ for the state $|l,n\rangle $ , where $%
l\geq 2$. It says that the states $|l,n\rangle $ ($l\geq 2$) are not in the $%
|GHZ\rangle $ class.

Remark 2. $\tau (\psi )$ in Eq. (\ref{odd-residua-def}) is considered as the
residual entanglement for odd $n$\ qubits in \cite{LDF07a} and \cite{LDF07e}%
. From the above discussion, we can say that for odd $n$ qubits, $%
|GHZ\rangle $ possess the maximal residual entanglement $\tau =1$ while the
residual entanglement $\tau $ for the Dicke states $|l,n\rangle $ vanishes.

\subsection{All the $n$-qubit symmetric Dicke states with $2\leq l\leq (n-2)$
excitations are different from the $n$-qubit $|W\rangle $ state under SLOCC.}

Lemma 3 in Appendix A says that $D^{(l)}(\psi )=0$ for any state $|\psi
\rangle $ in the $|W\rangle $ class. By Lemma 4 in Appendix A, $%
D^{(l)}(|l,n\rangle $ $)\neq 0$ for states $|l,n\rangle $ , where $2\leq
l\leq (n-1)/2$. Hence, states $|l,n\rangle $ ($2\leq l\leq (n-1)/2$) are not
in the $|W\rangle $ class.

Conjecture. For odd $n$ qubits, $|2,n\rangle $, $|3,n\rangle $, ..., and $%
|(n-1)/2,n\rangle $ are different from each other under SLOCC because $%
D^{(k)}(|l,n\rangle $ $)=0$ whenever $k\neq l$\ while $D^{(l)}(|l,n\rangle $ 
$)\neq 0$.

\section{Monogamy inequality for the $n$-qubit symmetric Dicke states}

Osborne and Verstraete in \cite{Osborne} obtained the general Coffman, Kundu
and Wootters monogamy inequality \cite{Coffman}. The general inequality is $%
C_{12}^{2}+...+C_{1n}^{2}\leq C_{1(2...n)}^{2}$. Let $\chi
_{12...n}=C_{1(2...n)}^{2}-(C_{12}^{2}+...+C_{1n}^{2})$. For the $n$-qubit $%
|GHZ\rangle $ state, $\chi _{12...n}=1$ \cite{Ou}, For the $n$-qubit $%
|W\rangle $ state, $\chi _{12...n}=0$ \cite{Coffman}. We show that all the $%
n $-qubit symmetric Dicke states but the $n$-qubit $|W\rangle $ state
satisfy $0<\chi _{12...n}<1$ below. For this purpose, first we need to
compute the concurrence between any two qubits and the one between qubit 1
and other qubits as follows.

\subsection{Concurrence between any two qubits}

Let $\rho _{12...n}=|l,n\rangle $ $\langle l,n|$. When any $(n-2)$ qubits
are traced out, let $\rho _{ij}$\ be the density matrices of the remaining
two qubits. For example,

$\rho _{12}=tr_{3..n}\rho _{12...n}=$

$(\frac{(n-l)(n-l-1)}{n(n-1)}|00\rangle \langle 00|+\frac{l(l-1)}{n(n-1)}%
|11\rangle \langle 11|)+\frac{2l(n-l)}{n(n-1)}|\psi ^{+}\rangle \langle \psi
^{+}|$, where $|\psi ^{+}\rangle =(|01\rangle +|10\rangle )/\sqrt{2}$. Note
that $|\psi ^{+}\rangle $ is a maximally entangled state of two qubits. By
the symmetry of the $n$-qubit symmetric Dicke states, all the reduced
density operators $\rho _{ij}=\rho _{12}$. By Coffman, Kundu and Wootters'
definition \cite{Coffman}, $\overline{\rho _{12}}=(\sigma _{y}\otimes \sigma
_{y})\rho _{12}(\sigma _{y}\otimes \sigma _{y})$. Then $\rho _{12}\overline{%
\rho _{12}}$ has the following eigenvalues: $\ \frac{%
4l^{4}-8l^{3}n+4l^{2}n^{2}}{n^{4}-2n^{3}+n^{2}}$, $0$, $\frac{1}{%
n^{4}-2n^{3}+n^{2}}\left(
l^{4}-2l^{3}n+l^{2}n^{2}+l^{2}n-l^{2}-ln^{2}+ln\right) $ (double roots).
Thus, the concurrence of the density matrix $\rho _{12}$ is

\begin{equation}
C_{12}=2\frac{\sqrt{l(n-l)}(\sqrt{l(n-l)}-\sqrt{\left( l-1\right) (n-l-1)})}{%
n(n-1)}\text{.}  \label{concurrence-1}
\end{equation}

For the definition of the two qubit concurrence, see \cite{Hill}. By
symmetry of the Dicke states, when any $(n-2)$ qubits are traced out the
concurrence between the remaining two qubits is $C_{12}$.

The concurrence shows that the $n$-qubit symmetric Dicke states\ retain
bipartite entanglement, even distillable. It is not hard to derive that the
concurrence decreases as the number of excitations increases. Therefore,
among the $n$-qubit symmetric Dicke states, the $n$-qubit $|W\rangle $ state
has the maximal concurrence: $2/n$. It has also been proven in \cite{Koashi}%
\ that the $|W\rangle $\ state optimizes the concurrence if all but two
qubits are traced out. In \cite{Koashi}\cite{Dur}, the concurrence of the $n$%
-qubit $|W\rangle $ state was also determined to be $2/n$. Among the even $n$%
-qubit symmetric Dicke states, the state $|n/2,n\rangle $\ possesses minimal
concurrence: $1/(n-1)$. In \cite{Stockton}, Stockton et al. also derived the
result. Among the odd $n$-qubit symmetric Dicke states, the state with $%
(n-1)/2$ excitations possesses the minimal concurrence: $\frac{(n+1)-\sqrt{%
(n+1)(n-3)}}{2n}\allowbreak $. While for the state $|GHZ\rangle $ of any $n$
qubits, when any $(n-2)$ qubits are traced out the concurrence between the
remaining two qubits vanishes, i.e., the remaining state is not entangled.

\subsection{Concurrence between one qubit and other $(n-1)$\ qubits}

By calculating,

\ \ \ $\rho _{1}=tr_{2..n}\rho _{12...n}=[\left( _{l}^{(n-1)}\right)
|0\rangle \langle 0|+\left( _{(l-1)}^{(n-1)}\right) |1\rangle \langle
1|]/\left( _{l}^{n}\right) =\frac{n-l}{n}|0\rangle \langle 0|+\frac{l}{n}%
|1\rangle \langle 1|$. By the definition in \cite{Coffman}, the concurrence
between qubit 1 and other qubits is

\begin{equation}
C_{1(2...n)}^{2}=4\det (\rho _{1})=4\frac{l(n-l)}{n^{2}}\text{.}
\label{concurrence-2}
\end{equation}

This concurrence demonstrates that the $n$-qubit symmetric Dicke states\
retain the entanglement between one qubit and other $(n-1)$\ qubits. It is
easy to see that the concurrence $C_{1(2...n)}^{2}$ increases as the number
of excitations does. Hence, the $n$-qubit $|W\rangle $ state possesses the
minimal concurrence between one qubit and other $n-1$\ qubits, i.e., $%
C_{1(2...n)}^{2}=$ $4\frac{n-1}{n^{2}}$. The even $n$-qubit symmetric Dicke
state $|n/2,n\rangle $\ possesses the maximal concurrence between one qubit
and other $(n-1)$\ qubits, i.e., $C_{1(2...n)}^{2}=1$. Among the odd $n$%
-qubit symmetric Dicke states, the state with $(n-1)/2$ excitations
possesses the maximal concurrence$\allowbreak $ between one qubit and other $%
(n-1)$\ qubits, i.e., $C_{1(2...n)}^{2}=\frac{n^{2}-1}{n^{2}}$. So far, we
only know that the $|GHZ\rangle $ state of any $n$ qubits has the maximal
concurrence$\allowbreak $ between one qubit and other $(n-1)$\ qubits, i.e., 
$C_{1(2...n)}^{2}=1$.

\subsection{Monogamy inequality for Dicke states}

By the symmetry of the $n$-qubit symmetric Dicke states and from Eq. (\ref%
{concurrence-1}), $C_{12}^{2}+...+C_{1n}^{2}=4\frac{l(n-l)(\sqrt{l(n-l)}-%
\sqrt{\left( l-1\right) (n-l-1)})^{2}}{n^{2}(n-1)}$. Then from Eq. (\ref%
{concurrence-2}), $\chi _{12...n}=8l(n-l)\frac{\sqrt{-n+ln-l^{2}+1}(\sqrt{%
ln-l^{2}}-\sqrt{-n+ln-l^{2}+1})}{n^{2}\left( n-1\right) }$. Clearly, $0\leq
\chi _{12...n}<1$. Especially, for the $|W\rangle $ state $\chi _{12...n}=0$%
. Thus, it verifies Eq. (27) in \cite{Coffman}. When $l<n/2$, $\chi
_{12...n} $ increases as $l$ does. For even $n$ qubits,\ when $l=n/2$, $\chi
_{12...n}$ gets the maximum $\frac{n-2}{n-1}$. For odd $n$ qubits,\ when $%
l=(n-1)/2$, $\chi _{12...n}$ gets the maximum $\frac{(n+1)(n-1)\sqrt{n-3}[%
\sqrt{(n+1)}-\sqrt{n-3}]}{2n^{2}}<1$. When $l>n/2$, $\chi _{12...n}$
decreases as $l$ increases. Therefore, when $2\leq l\leq (n-2)$, $0<\chi
_{12...n}<1$. Clearly, $\chi _{12...n}$ is almost 1 for large $n$ when $l$
is about $n/2$.

For example, when $n=5$ and $l=2$, $\chi _{12345}=\allowbreak 0.702\,77$.
When $n=6$ and $l=2$, $\chi _{123456}=\allowbreak 0.675\,19$. When $n=4$ and 
$l=2$,$\ \chi _{1234}=2/3$. When $n=6$ and $l=3$, $\chi _{123456}=4/5$.

$\allowbreak $

\section{Summary}

\ In this paper, we show that the $n$-qubit symmetric Dicke states with $l$ (%
$2\leq l\leq (n-2)$) excitations are different from the states $|GHZ\rangle $
and $|W\rangle $ of $n$ qubits under SLOCC, respectively. We also argue that
the even $n$-qubit symmetric Dicke state with $n/2$ excitations is different
from the even $n$-qubit symmetric Dicke states with $l\neq n/2$ excitations
under SLOCC. And we demonstrate that all the $n$-qubit symmetric\ Dicke
states with $l$ ($2\leq l\leq (n-2)$) excitations satisfy Coffman, Kundu and
Wootters' generalized monogamy inequality $%
C_{12}^{2}+...+C_{1n}^{2}<C_{1(2...n)}^{2}<1$. We indicate that among the $n$%
-qubit symmetric Dicke states, the $n$-qubit $|W\rangle $ state maximally
retains the concurrence between the remaining two qubits when $(n-2)$ qubits
are traced out\ but possesses the minimal concurrence between one qubit and
other $n-1$\ qubits, while among the even $n$-qubit symmetric Dicke states,
the state $|n/2,n\rangle $ minimally retains the concurrence between the
remaining two qubits when $(n-2)$ qubits are traced out but possesses the
maximal concurrence between one qubit and other $(n-1)$\ qubits, i.e., $%
C_{1(2...n)}^{2}=1$.

\section*{Appendix A: The properties of $D^{(l)}(\protect\psi )$}

\setcounter{equation}{0} \renewcommand{\theequation}{A\arabic{equation}}

Lemma 1. The complementary states are SLOCC equivalent

Let the set of the basis states of $n$ qubits be $B=\{|0\rangle $, $%
|1\rangle $, ... , $|2^{n}-1\rangle \}$. Let $\bar{1}$ ( $\bar{0}$ ) be the
complement of a bit 1 $(0)$. Then $\bar{0}$ $=1$ and $\bar{1}=0$. Let $\bar{z%
}=\bar{z_{1}}\bar{z_{2}}...\bar{z_{n}}$ denote the complement of a binary
string $z=z_{1}z_{2}....z_{n}$. Also, the set of the basis states $B=\{|\bar{%
0}\rangle ,|\bar{1}\rangle ,...,|\overline{2^{n}-1}\rangle \}$. Let $%
|\varphi \rangle $ be any state of $n$ qubits. Then we can write $|\varphi
\rangle =$ $c_{0}|0\rangle $ $+c_{1}|1\rangle $ $+....+c_{2^{n}-1}|(2^{n}-1)%
\rangle $. Let $|\overline{\varphi }\rangle =c_{0}|\bar{0}\rangle $ $+c_{1}|%
\bar{1}\rangle $ $+....+c_{2^{n}-1}|\overline{(2^{n}-1)}\rangle $. We call $|%
\overline{\varphi }\rangle $ the complement of $|\varphi \rangle $.

Let $\sigma _{x}=\left( 
\begin{array}{cc}
0 & 1 \\ 
1 & 0%
\end{array}%
\right) $. Then $\sigma _{x}\otimes ...\otimes \sigma _{x}|\varphi \rangle
=\sum_{i=0}^{2^{n}-1}c_{i}(\sigma _{x}\otimes ...\otimes \sigma
_{x}|i\rangle )=\sum_{i=0}^{2^{n}-1}$ $c_{i}|\bar{\imath}\rangle =|\overline{%
\varphi }\rangle $.

Consequently, if two states of $n$ qubits are complementary then they are
SLOCC equivalent.

Lemma 2.

Let $|\psi \rangle $ be any pure state of $n$ qubits. Then, we can write

\begin{equation*}
|\psi \rangle =\sum_{i=0}^{2^{n}-1}a_{i}|i\rangle .
\end{equation*}%
Then, if $|\psi \rangle $ is equivalent to $|GHZ\rangle $ under SLOCC then $%
D^{(l)}(\psi )=0$. Especially, $D^{(l)}(GHZ)=0$.

Proof. It is known that $|\psi \rangle $ is equivalent to $|GHZ\rangle $
under SLOCC if and only if there exist invertible local operators $F^{(1)}$, 
$F^{(2)}$,..., $F^{(n)}$, where $F^{(i)}=\left( 
\begin{array}{cc}
f_{1}^{(i)} & f_{2}^{(i)} \\ 
f_{3}^{(i)} & f_{4}^{(i)}%
\end{array}%
\right) $, such that 
\begin{equation}
|\psi \rangle =\underbrace{F^{(1)}\otimes F^{(2)}\otimes ...\otimes F^{(n)}}%
_{n}|GHZ\rangle .  \label{eqcond}
\end{equation}%
Let $i_{n}i_{n-1}...i_{2}i_{1}$ be the $n$-bit binary number of $i$, where $%
i_{j}\in \{0,1\}$. By solving Eq. (\ref{eqcond}),

\begin{equation}
a_{i}=(f_{(2i_{n}+1)}^{(n)}...f_{(2i_{m}+1)}^{(m)}...f_{(2i_{1}+1)}^{(1)}+f_{(2i_{n}+2)}^{(n)}...f_{(2i_{m}+2)}^{(m)}...f_{(2i_{1}+2)}^{(1)})/%
\sqrt{2},  \label{amplitudes}
\end{equation}%
$i=0,1,...,2^{n}-1$. By substituting $a_{i}$ in Eq. (\ref{amplitudes}) into $%
D^{(l)}(\psi )$ in Eq. (\ref{D-cretiria}), \ one can verify that $%
D^{(l)}(\psi )=0$, where $l\geq 2$. It implies that $D^{(l)}(\psi )=0$ for
any state $|\psi \rangle $ in $|GHZ\rangle $ class.

Lemma 3. Let $|\psi \rangle $ be any pure state of $n$ qubits. Then, if $%
|\psi \rangle $ is equivalent to $|W\rangle $ under SLOCC then $D^{(l)}(\psi
)=0$. Especially, $D^{(l)}(W)=0$.

Proof. It is known that $|\psi \rangle $ is equivalent to $|W\rangle $ under
SLOCC if and only if there exist invertible local operators $F^{(1)}$, $%
F^{(2)}$,..., $F^{(n)}$, such that 
\begin{equation}
|\psi \rangle =\underbrace{F^{(1)}\otimes F^{(2)}\otimes ...\otimes F^{(n)}}%
_{n}|W\rangle .  \label{eq-to-w}
\end{equation}%
Let%
\begin{equation*}
\overline{\mathcal{\delta }}(i,j)=\left\{ 
\begin{array}{rcc}
1 & : & i=j, \\ 
0 & : & \text{otherwise.}%
\end{array}%
\right.
\end{equation*}%
And let $i_{n}i_{n-1}...i_{2}i_{1}$ be the $n$-bit binary number of $i$,
where $i_{j}\in \{0,1\}$.\ By solving Eq. (\ref{eq-to-w}),

\begin{equation}
a_{i}=(\sum_{j=n}^{1}\prod_{k=1}^{n}f_{2i_{k}+2^{\overline{\mathcal{\delta }}%
(j,k)}}^{(k)})/\sqrt{n},  \label{W-amplitudes}
\end{equation}%
$i=0,1,...,2^{n}-1$. By substituting $a_{i}$ in Eq. (\ref{W-amplitudes})
into $D^{(l)}(\psi )$ in Eq. (\ref{D-cretiria}),\ one can verify that $%
D^{(l)}(\psi )=0$, where $l\geq 2$. It implies that $D^{(l)}(\psi )=0$ for
any state $|\psi \rangle $ in the $|W\rangle $ class. Especially, $%
D^{(l)}(W)=0$.

Lemma 4. $D^{(l)}(|l,n\rangle $ $)\neq 0$ for the state $|l,n\rangle $ ,
where $2\leq l\leq (n-2)$.

Proof. For the state $|l,n\rangle $ , $a_{3+\Delta }a_{6+\Delta
}=a_{9+\Delta }a_{12+\Delta }=1/\left( _{l}^{n}\right) $ because $N(3+\Delta
)=N(6+\Delta )=N(9+\Delta )=N(12+\Delta )=l$, and $a_{1+\Delta }a_{4+\Delta
}=a_{0+\Delta }a_{5+\Delta }=a_{11+\Delta }a_{14+\Delta }=a_{10+\Delta
}a_{15+\Delta }=$ $a_{2+\Delta }a_{7+\Delta }=a_{8+\Delta }a_{13+\Delta }=0$
by the amplitudes of the state $|l,n\rangle $ . A simple calculation shows $%
D^{(l)}(|l,n\rangle $ $)\neq 0$. Especially when $n$ is even and $l=n/2$, $%
D^{(n/2)}(|n/2,n\rangle )\neq 0$.

\end{document}